\documentstyle[kaspp,twoside]{book}

\newcommand{\hmpc}{\;h^{-1}{\rm Mpc}}
\newcommand{\degree}{^{\circ}}
\newcommand{\degreez}{~\! . \!\! ^{\circ} }

\makehead{PARK AND LEE}{CORRELATION FUNCTIONS OF THE APM CLUSTERS}

\begin{document}
\jkastitle{31}{0}{0}{1998}

\title{\hspace{2.6cm} CORRELATION FUNCTIONS 
\newline OF THE APM CLUSTERS OF GALAXIES}

{\center
{\large\textbf{\textsc{C}}}\textbf{\textsc{HANGBOM}}
{\large\textbf{\textsc{P}}}\textbf{\textsc{ARK}}
\textbf{\textsc{AND}}
{\large\textbf{\textsc{S}}}\textbf{\textsc{UNGHO}}
{\large\textbf{\textsc{L}}}\textbf{\textsc{EE}}
\endcenter}
\vspace{-0.2cm}

\affil{Department of Astronomy, Seoul National University, Seoul 151-742, Korea
\\
E-mail: cbp@astro.snu.ac.kr, leesh@astro.snu.ac.kr}

\vspace{0.6cm}
\begin{abstract}
We have found that the two-point correlation function of
the APM clusters of galaxies has an amplitude much higher
than that claimed by the APM group.
As the richness limit increases from ${\cal R} = 53$ to 80,
the correlation length increases from 17.5 to $28.9 \hmpc$.
This indicates that the richness dependence of the APM cluster
correlation function is also much stronger than what the
APM group has reported.
The richness dependence can be represented by a fitting formula
$r_0 = 0.53 d_c + 0.01$, which is consistent with 
the Bahcall's formula $r_0 = 0.4 d_c$.
We have tried to find the possible reason for discrepancies.
However, our estimates for the APM cluster correlation
function are found to be robust against variation of the
method of calculation and of sample definition.
\end{abstract}

\keywords {cosmology: large-scale structure of universe --- galaxies: clusters of --- galaxies: clustering}

\section{INTRODUCTION}

Rich clusters of galaxies have been used for studies of large-scale
structure on scales $\simeq 10 \sim 100 \hmpc$. The spatial two-point
correlation function for the Abell clusters has been estimated by many authors
(see the review by Bahcall 1988; also Huchra et al. 1990;
Postman et al. 1992; Peacock \& West 1992),
which has been found to be consistent in shape with the power law form
\begin{equation}
\xi_{cc}(r) = \left( {r \over r_0} \right)^{-\gamma}
\end{equation}
with the correlation length $r_0 \simeq 20 \sim 26 \hmpc$
and with the power law index $\gamma \simeq 2$.
More recently new catalogs of clusters have been obtained by
automated selecting algorithm from the Edinburgh-Durham Southern Galaxy Catalog
(Lumsden et al. 1992), and from the APM Galaxy Survey (Dalton et al. 1997).
Redshift surveys of these new clusters
(Collins et al. 1995; Dalton et al. $1994a$) have been used to
estimate the spatial two-point correlation function (Nichol et al. 1992;
Dalton et al. $1994b$). The correlation length $r_0$
measured for these new clusters is reported to be lower than
that for the Abell clusters, that is
$14 \lesssim r_0 \lesssim 16 \hmpc$.

It has been argued that the counting radius
($r_c = 1.5 \hmpc$)
for the Abell clusters is so large that
the catalog contains projection effects,
which cause artificial line-of-sight correlations (Sutherland 1988;
Efstathiou et al. 1992). In addition to this, it has been recognized that
the intrinsically subjective nature of the Abell catalog causes
problems in homogeneity and statistical completeness (Postman et al. 1986).
However, Bahcall \& West (1992) have suggested that the
discrepancy between the Abell clusters and the APM clusters can be accounted by
the richness dependence of cluster correlation amplitudes (Bahcall 1988;
Bahcall \& West 1992)
\begin{equation}
r_0 = 0.4 d_c ,
\end{equation}
where $d_c$ is the mean intercluster separation
(Here $d_c = n_c ^{-1/3}$ ,where $n_c$ is the mean space density of clusters).
But Croft et al. (1997) have analyzed richness subsamples
of the APM clusters and argued that there is only
a weak dependence of correlation
amplitude with cluster richness, and that this disagrees with results
from the Abell clusters.

We describe our statistical richness subsamples drawn from
the APM cluster catalog in Section II. 
The method of calculation of the spatial two-point correlation function
and our results are presented in Section III. 
In Section IV we discuss many possibilities which might affect
the estimate of the correlation function.

\section{THE APM CLUSTER CATALOG}

Dalton et al.(1997) have published the catalog of
APM clusters which is complete over richness range
${\cal R} \ge 40$ and the characteristic magnitude range
$17.5 \le m_x \le 19.4$ (the distance estimation $m_x$ corresponds to
Abell's $m_{10}$ ). The catalog covers the region of 
the sky $21^{h} \lesssim \alpha \lesssim 5^{h}$ and
$-72 \degreez 5 \lesssim \delta \lesssim -17 \degreez 5$
of the APM Galaxy Survey (Maddox et al. 1990;
see also Loveday et al. 1996) and contains
957 clusters (see Fig.~1).
The exact APM survey field definition can be found in
the Astronomical Data Center (http://adc.gsfc.nasa.gov/).
Fig. 1 shows the distribution of 957 APM clusters (open circles),
185 APM fields (open squares), and 1456 holes (filled squares)
in the ($\alpha$, $\delta$) plane (Loveday et al. 1996).
Holes are the regions contaminated by big bright objects.

Although the 957 sample is complete
in itself, it contains only 374 clusters with measured
redshift. In the range of ${\cal R} \ge 53$ and $17.5 \le m_x \le 19.2$,
however, the completeness of the redshift sample is 90.9\% and 
increases up to 100\% at higher richness limits. 
So we generate four statistical subsamples with ${\cal R} \ge 53$, 
${\cal R} \ge 60$, ${\cal R} \ge 70$, and ${\cal R} \ge 80$, 
which contains 213, 149, 76, and 43 
clusters, respectively.
We have further restricted our APM samples to the declination range
$-65 \degree \le \delta \le -25 \degree$ to reduce the strong
dependence of number of clusters in declination (see below).
To convert the redshifts to comoving distances we assumed
the Einstein-de Sitter universe with $\Omega_0 = 1$.

\section{CLUSTER CORRELATION FUNCTION}

We estimate the spatial two-point correlation functions 
using the Hamilton's estimator (1993)
\begin{equation}
\xi_{cc}(r) = {{(DD)(RR)} \over {(DR)^{2}}}
{{4N_{c}N_{r}} \over {(N_{c} - 1)(N_{r} - 1)}} - 1 ~,
\end{equation}
which is evaluated to be less affected by uncertainties in the selection
function for $\xi_{cc} < 1$ (Dalton et al. $1994b$).
In equation (3), DD is the number of pairs in the sample with $N_c$ clusters, 
RR is the number of pairs in a random sample with $N_r$ random points,
and DR is the cluster-random pair count. 
The numerical factor 4 in the normalization term of pair
counts accounts for the fact that we count each DD or RR pair only once.
The uncertainties of $\xi_{cc}$ are computed from 
the simple equation $\delta \xi_{cc} = (1 + \xi_{cc})/\sqrt{DD}$,
which is easy to calculate but may underestimate the cosmic variance
in comparison to simulations (Croft \& Efstathiou 1994).

Croft et al. (1997) have estimated the correlation function
of the APM clusters located in the distance range of 
$50 \hmpc \le r \le 500 \hmpc$. They used a selection function
obtained by smoothing the distribution of clusters in redshift space
over $40 / \sqrt{2} \hmpc$ by a Gaussian filter (dashed lines in Fig.~~2).
The random catalog is generated in accordance with this radial 
selection function within the sample boundaries.
Equal weight is given to clusters and random points in the calculation
of correlation functions.

We have calculated the correlation function of the APM clusters
in the exactly same way that Croft et al. (1997) have followed.
Our estimates of correlation functions are plotted in Fig.$~~ \! 3$ for
four subsamples; the ${\cal R} \ge 53$ (filled dots),
${\cal R} \ge 60$ (triangles),
${\cal R} \ge 70$ (squares), and ${\cal R} \ge 80$ (open circles) samples.
The correlation lengths are $17.5^{+1.4}_{-1.5}$, $23.4^{+2.0}_{-2.0}$,
$25.1^{+3.6}_{-3.7}$, and $28.9^{+7.4}_{-7.1} \hmpc$, respectively.
For a comparison the fitting lines to Croft et al. (1997)'s
${\cal R} \ge 50$ and ${\cal R} \ge 70$ samples are drawn
as solid and dashed lines, respectively. 
It can be seen that there are clear discrepancies
between our results and Croft et al (1997)'s.
\begin{minipage}[t]{16.5cm}
   \vspace{-3cm}
   \hspace{-0.5cm}
   \epsfxsize=15cm
   \epsfbox{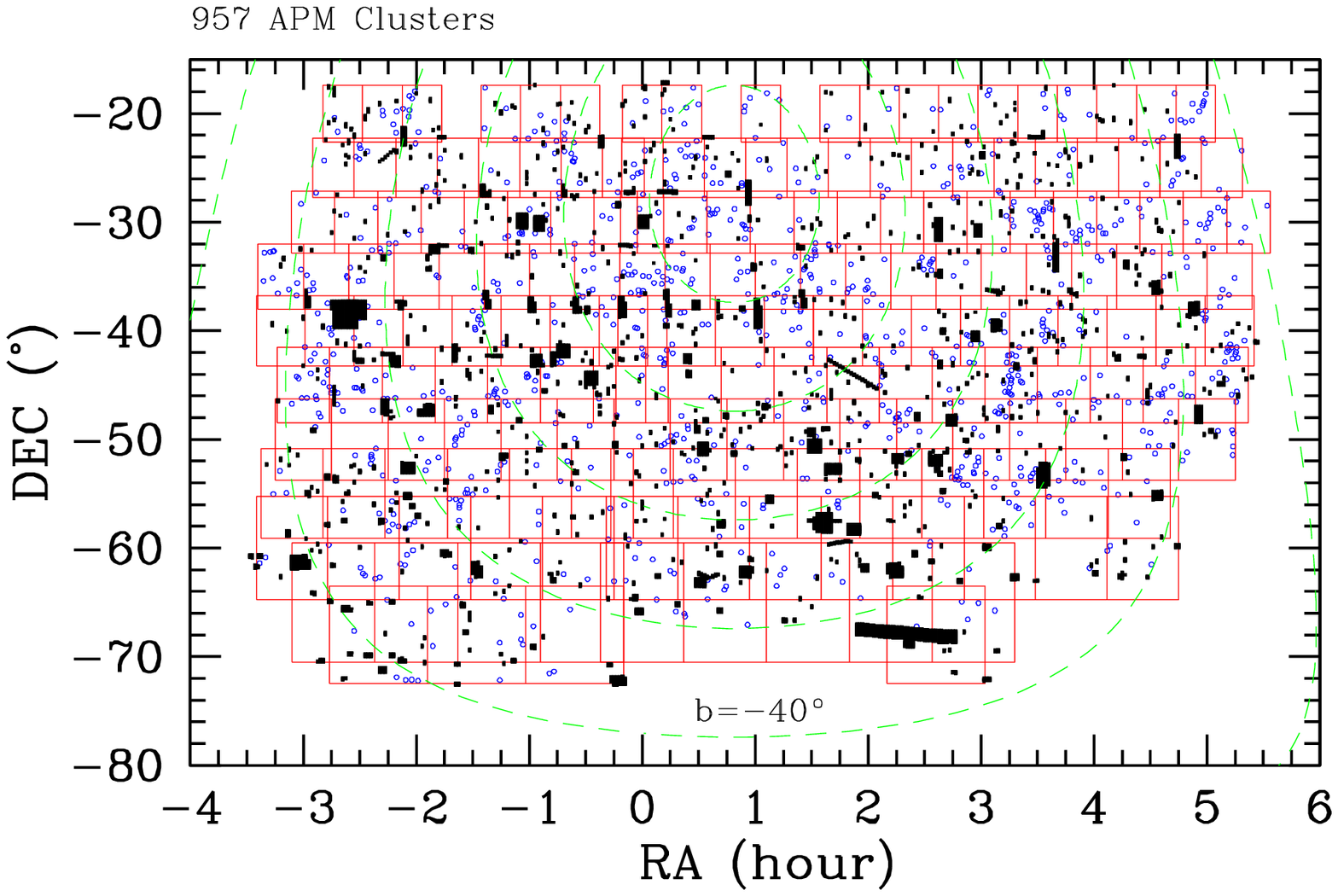}
\end{minipage}

\vspace{-3.2cm}
{\footnotesize
\hspace{-0.7cm}
\parbox[t]{17.5cm}{{\bf Fig. 1.}
The distribution of 957 APM clusters (open circles),
185 APM fields (open squares), and 1456 holes (filled squares).
The long dashed lines represent the galactic latitude.}}

\begin{minipage}[t]{16.5cm}
   \vspace{-1.3cm}
   \hspace{1.0cm}
   \epsfxsize=14cm
   \epsfbox{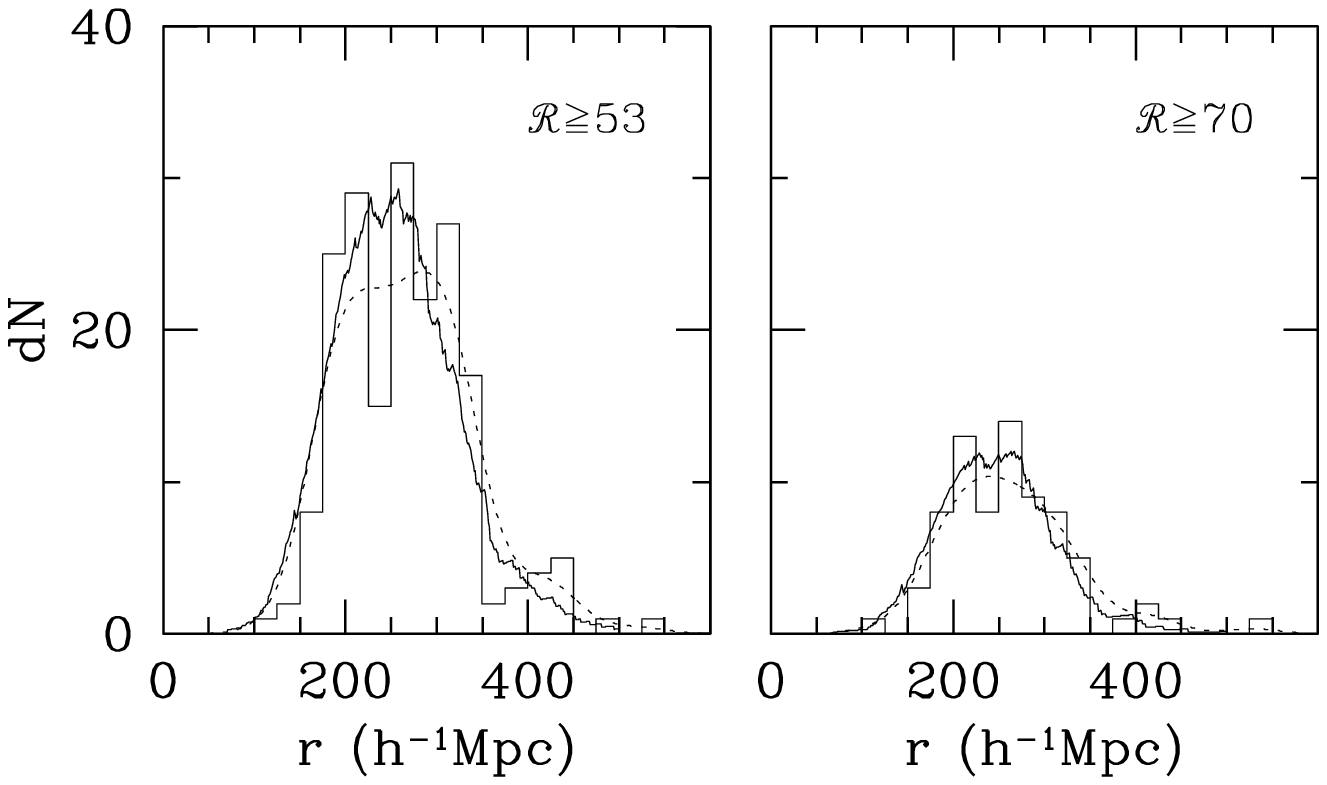}
\end{minipage}

\vspace{-6.5cm}
{\footnotesize
\hspace{-0.7cm}
\parbox[t]{17.5cm}{{\bf Fig. 2.}
Distributions of clusters with distance for subsamples with
${\cal R} \ge 53$ and ${\cal R} \ge 70$. 
Short dashed curves are distributions smoothed by a 
$40 / \sqrt{2} \hmpc$ Gaussian.
Solid lines are the selection function 
obtained by eq. (4). 
\vspace{1cm}}}
Amplitudes of correlation functions calculated by us are much
higher both for ${\cal R} \ge 53$ and ${\cal R} \ge 80$ samples. 
Furthermore, the richness dependence is much stronger.
The richness dependence of correlation functions of our APM subsamples 
can be described by a formula $r_0 = 0.53 d_c + 0.01$.
This richness dependence is stronger even than that reported for the
Abell clusters (Bahcall \& West 1992); $r_0 = 0.4 d_c$, quite contrary
to Croft et al. (1997)'s claim.

\section{DISCUSSION}

We have looked for various possibilities that could explain the
discrepancies between our and Croft et al. (1997)'s estimates of
the APM cluster correlation function.

\begin{minipage}[t]{16.5cm}
   \vspace{-1.5cm}
   \hspace{1.9cm}
   \epsfxsize=11cm
   \epsfbox{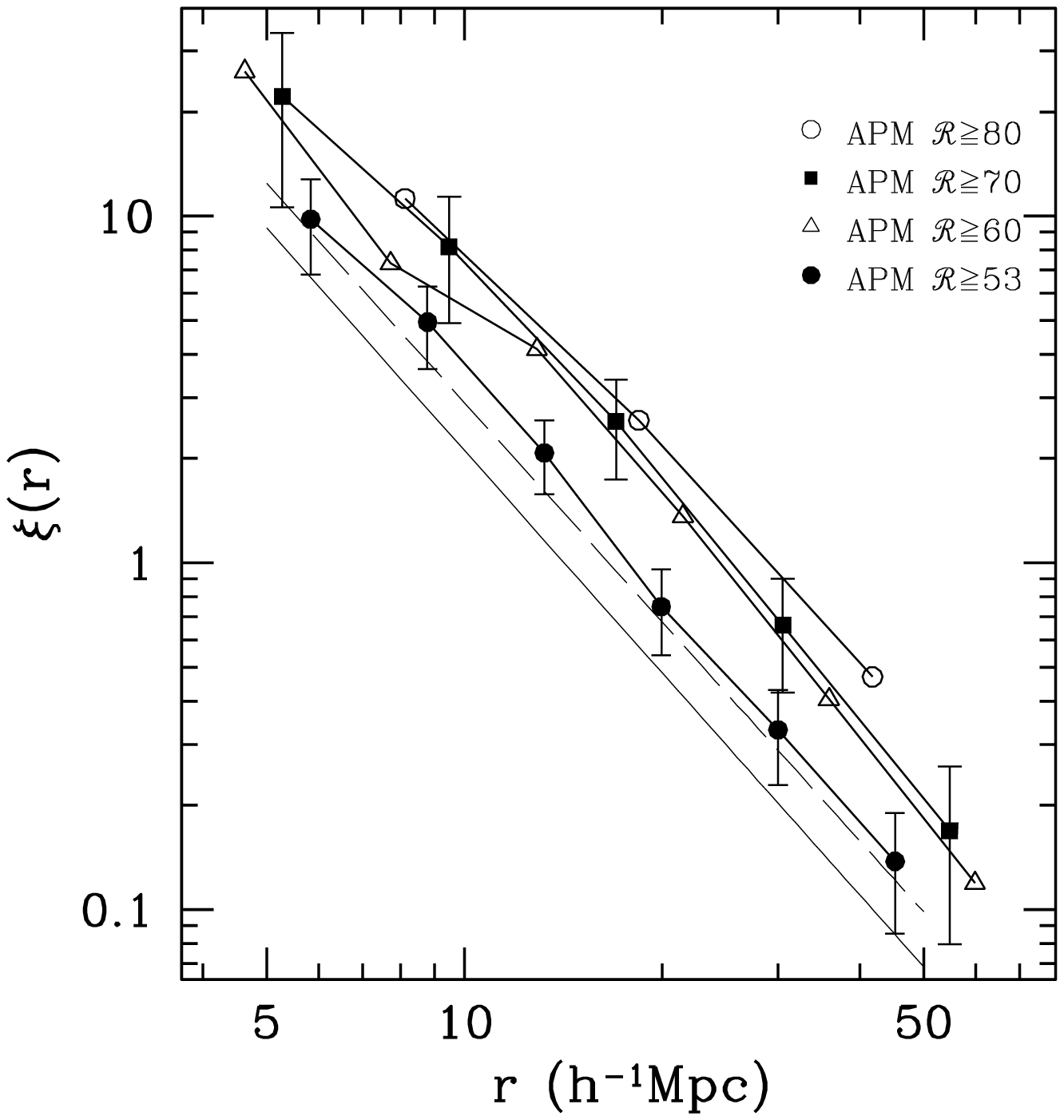}
\end{minipage}

\vspace{-1.5cm}
{\footnotesize
\hspace{-0.7cm}
\parbox[t]{17.5cm}{{\bf Fig. 3.}
Correlation functions of the APM cluster subsamples with
${\cal R} \ge 53$ (filled circles), ${\cal R} \ge 60$ (triangles), 
${\cal R} \ge 70$ (squares), and ${\cal R} \ge 80$ (open circles).
We also present two fitted lines of Croft et al. (1997)
for comparison; the solid line is for their sample B with ${\cal R} \ge 50$,
and the dashed line is for their ${\cal R} \ge 70$ sample. 
\vspace{1cm}}}

First, we changed the method calculating the selection function.
Instead of using the distribution of clusters in redshift space
smoothed over $40 / \sqrt{2} \hmpc$, we calculated the selection
function from the conventional formula
\begin{equation}
S(r) = \sum_{i; \atop {{D_{max,i} \ge r} \atop {D_{min,i} \le r}}}
{1 \over {{\Omega \over 3}
( D_{max,i}^{3} - D_{min,i}^{3} ) }} ~,
\end{equation}
where $\Omega$ is the solid angle of the survey region.
$D_{max,i}$ and $D_{min,i}$ are the maximum and minimum distances
to which the $i$-th cluster could be included in the sample
for the faint and bright magnitude limits of the survey
$m_{lim,upp} = 19.2$ and $m_{lim,low} = 17.5$, respectively.
The selection functions calculated by this formula are shown in Fig. 2
for the ${\cal R} \ge 53$ and ${\cal R} \ge 70$ samples. We have calculated the
correlation functions of the APM subsamples using these new selection
functions. However, there is practically no change in the amplitude
and in the richness dependence of the cluster correlation functions.

Second, we limited the distance range of clusters to 
$170 \hmpc < r < 330 \hmpc$ instead of $50 \hmpc < r < 500 \hmpc$
to eliminate the possible dominance of shot noises.
Again this did not affect our results.

Third, we changed the way to give weights to clusters and random points
in the calculation of correlation functions. Instead of giving equal
weights we gave weights equal to the inverse of the selection function.
To reduce the shot noise we limited the subsamples to the distance range
of $170 \hmpc < r < 330 \hmpc$.
This time, amplitudes of the correlation
functions became slightly higher than those shown in Fig. 3.
But this scheme still gives results which are inconsistent with
Croft et al. (1997)'s weak cluster correlation function.

Finally, we inspected the distribution of clusters in declination and
galactic latitude spaces. If the sample is affected by large galactic
obscuration or airmass variation across the sky, a selection function
varying on the sky should be taken into account.
Fig. 4 shows the number of clusters in declination (squares) or in
galactic latitude (circles) strips which have equal areas on the sky.
Points are located at the centers of strips. It can be shown that,
while there is no large fluctuation in the number of clusters
in the galactic latitude space, there is a rather strong variation
in the declination space for the ${\cal R} \ge 53$ and
${\cal R} \ge 70$ subsamples.
\begin{minipage}[t]{16.5cm}
   \vspace{-1.7cm}
   \hspace{1.8cm}
   \epsfxsize=13cm
   \epsfbox{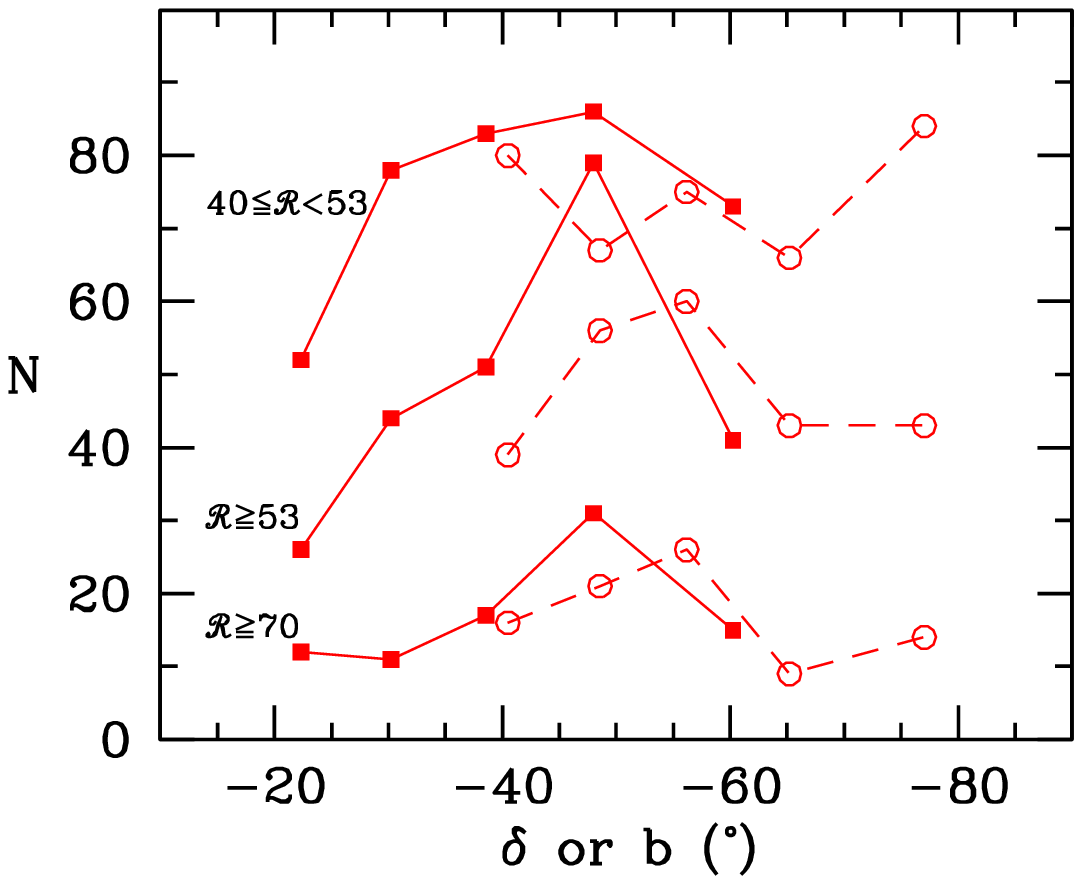}
\end{minipage}

\vspace{-5cm}
\hspace{1.3cm}
{\footnotesize
\parbox[t]{16.5cm}{{\bf Fig. 4.}
Variation of the number of clusters
with declination (squares) and galactic latitude (circles). \vspace{1cm}}}
Large number of clusters are concentrated in the declination strip
centered at $\delta = -48 \degree$.
We first varied the declination limits from the original range
$-72 \degreez 5 \lesssim \delta \lesssim -17 \degreez 5$
to narrower ranges in the subsample with ${\cal R} \ge 53$.
The correlation function at short separations did not change but
became somewhat steeper. However, the decrease in the amplitude
of the correlation function at large separations is small,
leaving the correlation function still within the $1 \sigma$ error limits.
We then used the smoothed distribution of clusters in declination
to generate the random catalog. This eliminates the effect of the strong
variation of number of clusters with declination.
The resulting correlation function has again somewhat lower amplitude
at large separations, but the changes are again within the error limits.

All these tests indicate that our estimate for the APM cluster
correlation function is robust.
It remains to be seen what has made APM group's results different
from ours.

\acknowledgments
This work was supported by the Basic Science Research Institute Program
No. BSRI-97-5408.

\end{document}